\documentclass[reprint,twocolumn,showpacs,superscriptaddress,longbibliography,email,floatfix,aps,prl]{revtex4-2}

\usepackage{header}
\usepackage{tikz_helper}

\begin{document}
\title{From Classical Stochastic to Monitored Quantum Dynamics: Dynamical Phase Coexistence in East Circuit Models}
\author{Marcel Cech}
\affiliation{Institut f\"ur Theoretische Physik and Center for Integrated Quantum Science and Technology, Universit\"at T\"ubingen, Auf der Morgenstelle 14, 72076 T\"ubingen, Germany}
\author{Johan du Buisson}
\affiliation{Institut f\"ur Theoretische Physik and Center for Integrated Quantum Science and Technology, Universit\"at T\"ubingen, Auf der Morgenstelle 14, 72076 T\"ubingen, Germany}
\author{Cecilia De Fazio}
\affiliation{Institut f\"ur Theoretische Physik and Center for Integrated Quantum Science and Technology, Universit\"at T\"ubingen, Auf der Morgenstelle 14, 72076 T\"ubingen, Germany}
\author{Federico Carollo}
\affiliation{Dipartimento di Fisica, Sapienza Università di Roma, Piazzale Aldo Moro 2, 00185 Rome, Italy}
\affiliation{Centre for Fluid and Complex Systems, Coventry University, Coventry, CV1 2TT, United Kingdom}
\author{Igor Lesanovsky}
\affiliation{Institut f\"ur Theoretische Physik and Center for Integrated Quantum Science and Technology, Universit\"at T\"ubingen, Auf der Morgenstelle 14, 72076 T\"ubingen, Germany}
\affiliation{School of Physics and Astronomy and Centre for the Mathematics and Theoretical Physics of Quantum Non-Equilibrium Systems, The University of Nottingham, Nottingham, NG7 2RD, United Kingdom}

\begin{abstract}
    Kinetically constrained models have been widely studied in the context of glass formers and non-equilibrium statistical mechanics. Although their simple local rules often result in structureless static properties, their dynamics exhibit intricate emergent phenomena. In this work, we investigate monitored quantum circuit models that interpolate between classical stochastic and unitary quantum dynamics. For any finite measurement strength, the measurement records provide an experimentally accessible probe of  the emergence of dynamical phases. By interpreting space-time resolved records as microstates of a fictitious 1+1D spin system, we employ thermodynamic concepts that allow us to investigate the dynamical coexistence between an active and inactive phase. We combine insights from classical stochastic dynamics and numerical simulations of monitored quantum dynamics to investigate different signatures of this dynamical phase coexistence as the measurement strength is varied. Our results shed light on the persistence of dynamical phase coexistence in the quantum regime, offering insights into future experimental studies of complex many-body dynamics in quantum simulators. 
\end{abstract}

\maketitle


\begin{figure}
    \centering
    \includegraphics{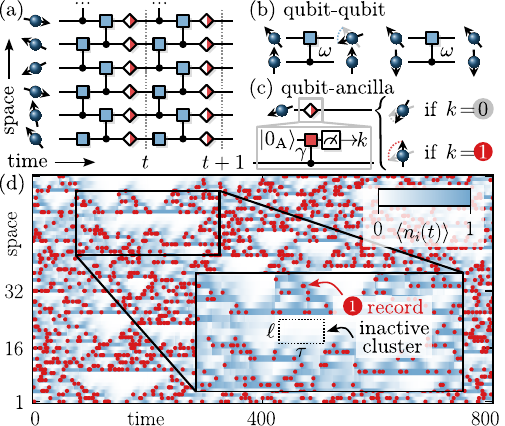}
    \caption{\textbf{Dynamical phase coexistence in monitored quantum East circuits.} (a)~The discrete-time dynamics consists of unitary gates in a brickwork structure (blue squares with control qubits below) and local monitoring (two-colored diamonds). (b)~Unitary gates implement qubit-qubit dynamics in terms of a conditioned rotation according to the East constraint if the control qubit is in state $\ket{1} \equiv \ket{\uparrow}$. (c)~Monitoring is achieved by ancilla-assisted operations that weakly measure the system according to Eq.~\eqref{eq:weak_measurements}. Their output $k_i(t) \in \{0, 1\}$ enters the so-called space-time measurement record $\eta$. (d)~Quantum expectation values $\expval{n_i(t)}$ of the occupation in state $\ket{1}$ and corresponding measurement history in a stochastic realization for $L=64$, $\omega = 0.1$ and $\gamma = 0.3$. Remarkably, we observe distinct active and inactive space-time clusters that are characterized by their respective number of $k = 1$ measurement outcomes. The dynamical phase coexistence can hence be accessed via the statistical properties of the time-integrated activity $A_{L, T} = \sum_{i,t} k_i(t)$ (cf.~Fig.~\ref{fig:fig3}), and of $\ell \times \tau$-sized inactive clusters (cf.~Fig.~\ref{fig:fig4}). }
    \label{fig:fig1}
\end{figure}

\textbf{Introduction.--- } The study of equilibrium phase transitions is of fundamental importance in the fields of statistical mechanics and condensed matter physics~\cite{Goldenfeld1992,Anderson1994,Ruelle2004,Chandler2005}. A hallmark of such phase transitions is an abrupt change in the macroscopic properties of a physical system when an external parameter, such as temperature or pressure, crosses a critical value. 

Interestingly, similar ideas can also be developed in an out-of-equilibrium setting. Here, \textit{dynamical} phase transitions exist, where properties of the dynamics change drastically while static ones remain unaffected~\cite{Garrahan2009}.
Paradigmatic examples of this phenomenon are portrayed by so-called kinetically constrained models~\cite{Ritort2003,Cancrini2008,Garrahan2010a}, where the (stochastic) dynamics is subject to local constraints that restrict on-site transitions depending on the configurations of neighboring degrees of freedom. Although these models are typically characterized by seemingly ``trivial'' static properties, e.g., structureless stationary states, they nevertheless display a variety of rich dynamical phenomena, including slow relaxation and glassy behavior~\cite{Sillescu1999,Garrahan2007,Hedges2009,Garrahan2009,Binder2011,Garrahan2018}, which can be studied in single dynamical realizations, or trajectories. A thermodynamically inspired analysis of these trajectories allows for the identification of dynamical phases by means of non-analytical behavior in their thermodynamic potentials~\cite{Touchette2009,Garrahan2010}. 

Although kinetically constrained models have been extensively studied in classical settings~\cite{Bodineau2012,Carollo2018,Banuls2019,Causer2020,Causer2021,Causer2022,Causer2022a,Casert2021,Sfairopoulos2025,Buca2019,Elmatad2013,Katira2018,Klobas2024,DeFazio2024}, advances in quantum simulation platforms have also attracted interest in their quantum counterparts~\cite{Bernien2017,GuardadoSanchez2020,Adler2024,Weckesser2024,Karch2025,Zhao2025a}. Beyond their unitary manifestations linked to observations of weak ergodicity breaking~\cite{Turner2018,Pancotti2020,Bertini2024,Badbaria2024,Menzler2025,BenAmi2025,Jonay2025}, an interesting avenue of research is the study of dynamical phase coexistence in single trajectories of open, i.e., monitored, quantum systems~\cite{Olmos2012,Lesanovsky2013a,Olmos2014,Horssen2015,Gribben2018,Rose2022,ValenciaTortora2022,Landi2023,Cech2025}. However, simulating their long-time dynamics is a daunting task
~\cite{Gillman2019,Fux2023,Hryniuk2024,Link2024,Boneberg2025,Westhoff2025,Marche2025,Sander2025} and only a few works~\cite{Causer2025,Cech2025a} have explored dynamical phase coexistence in large instances of such systems. Hence, a complete understanding of the fate of dynamical phase coexistence in the quantum regime is still lacking.

In this work, we address this question by investigating the dynamical phase coexistence as we move from classical to genuine quantum dynamics. To this end, we introduce and analyze a class of monitored quantum circuit models that interpolate between classical stochastic and unitary quantum dynamics. 
Here, the unitary brickwork circuit provided by the Floquet-quantum East model is combined with an ancilla-assisted measurement of all qubits at stroboscopic times [cf.~Fig.~\ref{fig:fig1}(a-c)]. 
These measurements give rise to an ensemble of stochastic realizations that are characterized by their respective space-time measurement records. While quantum expectation values of individual stochastic realizations are obscured by postselection overheads in potential experimental investigations, the space-time measurement records themselves can be accessed directly from single experimental runs. Consequently, we investigate their dynamical phase diagram and the statistics of clusters with identical measurement outcomes to characterize an emergent dynamical phase coexistence [cf.~Fig.~\ref{fig:fig1}(d)]. 
As a central result, we prove the existence of the underlying dynamical phase transition in the classical limit and demonstrate its persistence in the quantum regime. 
Moreover, our findings suggest that dynamical phase coexistence can be directly observed in quantum simulators that offer unitary control and mid-circuit measurements. \\


\textbf{Monitored quantum dynamics.---} We consider a system of $L$ qubits, each having states expressed in the computational basis $\{\ket{0},\ket{1}\}$. The discrete-time dynamics, illustrated in Fig.~\ref{fig:fig1}(a), consists of alternating layers of unitary gates (shown in blue) followed by
local monitoring of the excited state (shown in white-red in the same panel).
The unitary evolution is implemented locally on two qubits at positions $i\!-\!1$ and $i$ via the conditioned Quantum-East gate [cf.~Fig.~\ref{fig:fig1}(b)]
\begin{align}
    u_{i-1, i} = \EastGate = e^{-i \omega n_{i-1} \sigma_i^{x}} \, ,
    \label{eq:East_gate}
\end{align}
where $n = \ketbra{1}$ and $\sigma^x = \ketbra{0}{1} + \ketbra{1}{0}$. The projector $n_{i-1}$ enforces the East constraint~\cite{Jaeckle1991,Faggionato2013}, where the transition $\ket{0} \leftrightarrow \ket{1}$ of the $i$th qubit is only allowed if the preceding qubit has a non-zero overlap with the excited state $\ket{1}$. 
Note that we choose open boundary conditions with $n_0 = 1$ such that the first qubit is always driven.

To locally monitor the qubits at the end of every timestep, we employ the ancilla-assisted measurement shown in Fig.~\ref{fig:fig1}(c). Here, each qubit is coupled individually to its corresponding ancilla via a Quantum-East gate (in red) that is controlled by the parameter~$\gamma$. The qubit acts as a facilitator for the ancilla, which, at the start of every timestep, is in state $\ket{0_\mathrm{A}}$. Therefore, the projective measurement of the ancillas can produce outcomes $k = 1$ only if the corresponding qubits have non-zero overlap with $\ket{1}$. This (generally) non-projective measurement of the system qubits is described by the nonlinear action $\ket{\psi}\!~\!\WeakMeasurementDiamond\!~\!\ket{\psi'}$ in terms of the Kraus operators~\cite{Kraus1983,Breuer2002,Chen2024g}
\begin{equation}
    \label{eq:weak_measurements}
    \begin{aligned}
        K_{0, i} &=  \mathds{1} + (\cos\gamma - 1) \, n_i \, , \quad & \text{if $k_i = 0$\,,}\\ 
    K_{1, i} &= -i (\sin{\gamma})\, n_i \, , & \text{if $k_i = 1$\,.}
    \end{aligned}
\end{equation}
Here, $\ket{\psi'} = {K_{k, i} \ket{\psi}}/{\sqrt{\pi_k}}$ is the conditioned state for the system previously in state $\ket{\psi}$ upon observing the ancilla outcome $k_i$, with probability  $\pi_k = \| K_{k, i} \ket{\psi} \|^2$. Importantly, the parameter $\gamma$ controls the strength of the measurement. It allows us to interpolate between the classical stochastic Floquet-East model~\cite{DeFazio2024} (for projective measurements at $\gamma = \pi / 2$) and a purely unitary Floquet-Quantum East model~\cite{Bertini2024}~(for $\gamma = 0$), obtained by Trotterization of the Quantum East Model~\cite{Horssen2015,Pancotti2020}. 

Iterating the above procedure results in stochastic realizations, where the space-time resolved ancilla measurement outcomes $k_i(t)$ are collected in the trajectory $\eta = [k_i(t)]_{i, t}$. Figure~\ref{fig:fig1}(d) shows the $k = 1$ ancilla measurement records (red dots) superimposed on the expectation value of the excitation density $\expval{n_i(t)}$ (white to blue). We stress that ancilla records and the excitation density differ significantly in terms of experimental accessibility. This is because $\expval{n_i(t)}$ generally faces postselection overheads in repetitively preparing the same stochastic realization of the state $\ket{\psi(t)}$~\cite{Fisher2023}. In contrast, ancilla records are directly accessible from single experimental runs for any finite measurement strength $\gamma$. 
Importantly, we observe that the ancilla outcomes tend to reflect the system dynamics: the $k = 1$ outcomes concentrate (are absent) in regions where $\expval{n_i(t)}$ is large (small), which we refer to as active (inactive) clusters. Interestingly, we can view these clusters as instances of distinct dynamical phases coexisting within a single realization [cf.~Fig.~\ref{fig:fig1}(d)]. 
As such, these space-time records are a valuable tool for investigating the dynamical phase coexistence observed in Fig.~\ref{fig:fig1}(d) in terms of the presence of a dynamical phase transition akin to classical East models~\cite{Garrahan2007,Elmatad2013,Katira2018,Klobas2024,DeFazio2024}. \\


\textbf{Dynamical phase diagram.---} To characterize the different dynamical phases, we investigate the time-integrated activity $A_{L, T}(\eta) = \sum_{i,t} k_i(t)$ (where $t \in \{1, ..., T\}$ and $i \in \{1, ..., L\}$). We adopt this quantity as a dynamical order parameter and evaluate it on the space-time measurement records $\eta = [k_i(t)]_{i,t}$\,, which we interpret as microstates with probability $\pi(\eta)$, in analogy with statistical physics ~\cite{Garrahan2009,Cilluffo2021,Cech2023,Cech2025}. Using this perspective, we define a dynamical partition function $\mathcal{Z}_{L, T}(s) = \sum_\eta \pi(\eta) e^{-s A_{L, T}(\eta)}$, whose derivatives with respect to the counting field $s$, evaluated at $s = 0$, allow us to compute moments and cumulants of the activity. 
We focus here on the first scaled cumulant, i.e., the activity density,
\begin{align}
    a_{L, T}(s) = -\frac{1}{L T} \frac{\partial}{\partial s} \log \mathcal{Z}_{L, T}(s) = \frac{1}{LT} \expval{A_{L, T}(\eta)}_s  \,,
    \label{eq:activity_s_ensemble}
\end{align}
which, when evaluated at $s = 0$, yields the average activity density of the typical dynamics: $a_{L, T}(0) = \sin^2\!\gamma / 2$ (see Supplemental Material~\cite{SM}~\vphantom{\cite{Lesanovsky2013,Johansson2013,Fishman2022,Ferris2012}} for details)~\footnote{Note that, analogous to the expectation values in statistical mechanics, we have that $\expval{O(\eta)}_s = \sum_\eta O(\eta) \pi(\eta) e^{-s A_{L, T}(\eta)} / \mathcal{Z}_{L, T}(s)$.}. Note that the dynamical partition function $\mathcal{Z}_{L, T}(s)$ can in fact become non-analytic in the thermodynamic limit ($L,T \rightarrow \infty$). 
Particularly interesting is the case in which such singular behavior is found at $s=0$. For example, a discontinuous jump of the activity $a(s)$ indicates a first-order dynamical phase transition, which manifests in the dynamical coexistence of two phases with different activities in a single stochastic realization~\cite{Garrahan2007}.
When employing this perspective, the counting field $s$ plays the role of an inverse ``temperature''. Consequently, changing $s$ allows us to resolve the two phases: for $s>0$, the active phase dominates whereas, for $s < 0$, the inactive phase dominates.

\begin{figure}
    \centering
    \includegraphics{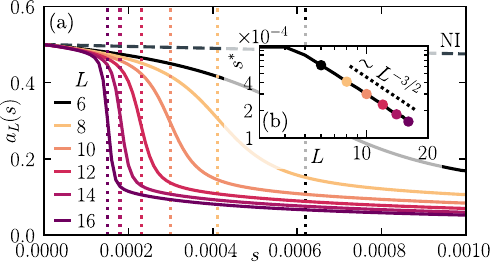}
    \caption{\textbf{Activity of the classical stochastic Floquet-East model} for $\gamma = \pi/2$ and $\omega = 0.1$.
    (a)~The activity density $a_L(s)$ is displayed for various system sizes $L$. As $L$ grows, we observe an increasingly sharp crossover (marked by dotted lines) between active and inactive phases. For comparison, we also show $a_L(s)$ for non-interacting (NI) spins, (indicated by the dashed line), where no sharp crossover exists. (b)~Inset showing the scaling of $s^* = \mathrm{argmax}_s\{a'(s)\}$ with system size $L$. The crossover $s^*$ approaches $0$ as the system size increases (see dotted line with $\sim L^{-3/2}$ as a guide for the eye). }
    \label{fig:fig2}
\end{figure}

To illustrate the idea, we now consider classical stochastic dynamics at $\gamma = \pi / 2$, where we can access exactly the limit of $T \to \infty$ for intermediate system sizes $L$. Additionally, we can analytically prove the existence of an active-inactive dynamical phase transition (see Supplemental Material~\cite{SM}).
In particular, we show that here the activity density $a(s) = \lim_{L, T \to \infty} a_{L, T}(s)$ displays a discontinuity at $s = 0$: we have $a(0) = 1/2$ and $a(s > 0) = 0$. This demonstrates that the typical dynamics at $s = 0$ lies on the coexistence line between the active and inactive phase. 
For finite $L$, where no strict non-analytic behavior can exist, Fig.~\ref{fig:fig2}(a) nevertheless shows that $a(s)$ has a sharp crossover between active and inactive phases at $s^* > 0$ [cf.~$s^* = \mathrm{argmax}_s\{a'(s)\}$], indicating that the active phase dominates in the typical dynamics. 
As $L$ increases, the crossover becomes increasingly sharp, and furthermore $s^*\rightarrow 0$ [cf.~Fig.~\ref{fig:fig2}(b)]. This means that the dynamical coexistence between active and inactive phases becomes increasingly evident in individual stochastic realizations. 
We thus can reasonably infer the presence of dynamical coexistence in the thermodynamic limit, even without the aforementioned proof~\cite{Garrahan2007,Bodineau2012,Causer2025}.

\begin{figure}
    \centering
    \includegraphics{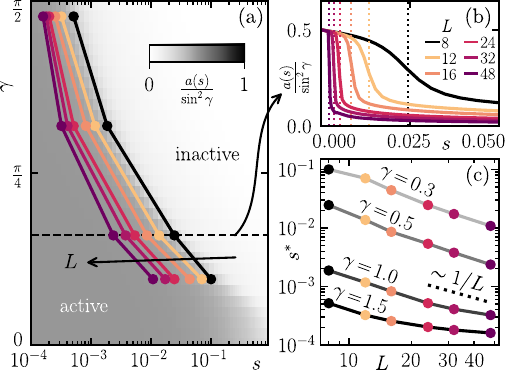}
    \caption{\textbf{Dynamical phase diagram.} (a)~Normalized activity density $a_{L, T}(s) / \sin^2\!\gamma$ for $L=8$, $\omega = 0.1$. As the counting field $s$ is varied, a sharp crossover emerges between active and inactive dynamical phases. The position of this crossover depends on the measurement strength~$\gamma$, but generally shifts toward $s = 0$ as the system size $L$ increases (colored symbols: $L=8$ to $L=48$). (b)~Cut at $\gamma = 0.5$. The crossover sharpens significantly with increasing $L$, approaching a discontinuity characteristic of a first-order phase transition. Dotted lines indicate the crossover positions $s^*$. (c)~Scaling analysis of the crossover. The crossover $s^*$ approaches $s = 0$ as $L$ increases (see dotted line $\sim 1/L$ for reference), hereby confirming the observation of dynamical phase coexistence in Fig.~\ref{fig:fig1}(d). }
    \label{fig:fig3}
\end{figure}

With these insights in mind, we now investigate the presence of analogous signatures in the monitored quantum dynamics away from $\gamma = \pi/2$. 
We refer to the Supplemental Material~\cite{SM} for details on the numerical methods that allow us to access large system sizes and long but finite times. In Fig.~\ref{fig:fig3}(a), we show the normalized activity density $a_{L, T}(s) / \sin^2\!\gamma$ for $L=8$ and $\omega = 0.1$ as a function of the counting field $s$ for different values of the measurement strength $\gamma$. Already for this system size, we observe two distinct phases that are separated by a sharp crossover. 
Interestingly, for fixed $\gamma$ [cf.~Fig.~\ref{fig:fig3}(b)], the behavior of the order parameter closely resembles that observed for the classical stochastic Floquet-East model in Fig.~\ref{fig:fig2}(a). 
However, for the monitored quantum dynamics, we find that the position of the crossover $s^*$ also depends on the measurement strength $\gamma$. In particular, increasingly rare realizations (associated with a larger magnitude of $s$~\cite{Carollo2018}) of the system dynamics are required to observe the crossover from the active to inactive phase.
Nevertheless, for larger system sizes $L$ (colored symbols representing $L=8$ to $L=48$), the crossover sharpens [cf.~Fig.~\ref{fig:fig3}(b)], and simultaneously shifts towards $s = 0$ for all considered values of $\gamma$ [cf.~Fig.~\ref{fig:fig3}(c)]. Combining these observations suggests clearly that the dynamical phase transition persists also away from $\gamma = \pi / 2$ in the thermodynamic limit. Note that the apparent flattening as $s^*$ moves closer to zero is a finite-time effect in the simulations (see Supplemental Material~\cite{SM} for a detailed discussion). \\


\begin{figure}
    \centering
    \includegraphics{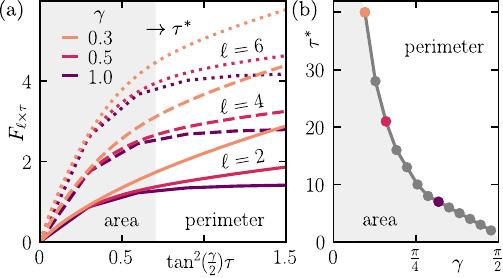}
    \caption{\textbf{Crossover in the statistics of inactive clusters. } (a)~Negative log-probability $F_{\ell \times \tau}=-\log p_{\ell \times \tau}$ of inactive $\ell \times \tau$ clusters in the ancilla records for fixed $\ell=2,4,6$ (solid, dashed, dotted) at $L=64$ and $\omega=0.1$. For small $\tau$, $F_{\ell \times \tau}$ grows proportionally to the area $\ell \tau$, consistent with nearly independent outcomes at average activity density $a_{L, T}(0)=\sin^2\!\gamma/2$. Around a $\gamma$-dependent crossover time $\tau^*$, the slope decreases and the growth becomes perimeter-like, signaling a collective nature of the inactive region. Rescaling time as $\tan^2(\gamma/2)\tau$ approximately collapses the crossover positions across $\gamma$ (see Supplemental Material~\cite{SM} for the perturbative derivation).
    (b)~Crossover time $\tau^*$ versus measurement strength $\gamma$. The crossover persists for all tested $\gamma$, indicating that active–inactive dynamical phase coexistence remains in the quantum regime. Notably, the unscaled $\tau^*$ grows as $\gamma$ decreases: this is because longer observation times are needed to separate collective inactive domains from inactivity caused only by lower average activity.}
    \label{fig:fig4}
\end{figure}

\textbf{Statistics of inactive clusters.---} We now investigate how precursors of the dynamical phase transition can be established at the level of individual trajectories. In particular, we focus on analyzing the statistics of extended inactive clusters in the ancilla records such as in  Fig.~\ref{fig:fig1}(d). To quantify this, we introduce the dynamical free energy~\cite{Katira2018,Klobas2024,DeFazio2024,Cech2025a} 
\begin{equation}
\label{eq:freeenergy}
    F_{\ell \times \tau}= -\log p_{\ell \times \tau}\,,
\end{equation}
where $p_{\ell \times \tau}$ is the probability of finding an inactive cluster of size $\ell \times \tau$ in the ancilla measurement records (see Supplemental Material~\cite{SM} for details). 
Figure~\ref{fig:fig4}(a) shows that the dynamical free energy exhibits a crossover between two distinct scaling regimes at a characteristic time $\tau^*$. For small clusters ($\tau < \tau^*$), $F_{\ell \times \tau}$ is proportional to the area $\ell\tau$, while for large clusters ($\tau \gg \tau^*$), it scales with the perimeter of the space-time region. While the area-scaling can also arise from uncorrelated measurement outcomes, perimeter-scaling results truly as a collective effect.
As we show in the Supplemental Material~\cite{SM}, this perimeter-scaling behavior of the dynamical free energy implies that, in the thermodynamic limit, the activity density in Eq.~\eqref{eq:activity_s_ensemble} vanishes on the positive branch $s > 0$.  Consequently, the transition point occurs at $s=0$ in the same limit. Indeed, the area-to-perimeter crossover is a characteristic pre-transition effect in other kinetically constrained models approaching a dynamical phase transition \cite{Katira2018,Klobas2024,DeFazio2024}, commonly termed a ``hydrophobic crossover''. This analogy originates in the change from volume- to surface-dominated free-energy scaling for increasingly large solutes in water near the liquid-vapor transition~\cite{Chandler2005}.

Furthermore, as shown in Fig.~\ref{fig:fig4}(b), the crossover time $\tau^*$ increases with decreasing measurement strength $\gamma$. 
This can be understood in the following way. As the measurement strength $\gamma$ decreases, also the average activity density $a_{L, T}(0) = \sin^2\!\gamma / 2$ decreases. 
Consequently, spurious large inactive clusters could be produced by uncorrelated ancilla measurement outcomes, and even larger inactive space-time clusters are required to infer the presence of correlated ancillas corresponding to a perimeter-scaling of the free energy. 
As a result, extrapolating to the $\gamma \to 0$ regime is a particularly challenging task.
In fact, this can also be seen in Fig.~\ref{fig:fig3}, where $s^*$ takes larger values as $\gamma$ decreases at a fixed system size $L$. Hence, larger system sizes are required to establish the dynamical phase coexistence. \\


\textbf{Outlook.---} Our work has established that dynamical phase coexistence can extend from classical dynamics ---where it was originally established in the study of glassiness--- to monitored quantum many-body systems. Furthermore, our investigation has identified the weak measurement regime as a particularly attractive direction for future study. Here, our results suggest that larger and larger system sizes, in both the spatial and temporal dimensions, are required to characterize this dynamical phase coexistence. As this ultimately exhausts classical resources, this direction seems particularly promising for investigation with quantum computers and simulators with mid-circuit readout capabilities. \\

The code and data supporting the findings of this work are available on Zenodo \cite{ZenodoData}. \\


\let\oldaddcontentsline\addcontentsline 
\renewcommand{\addcontentsline}[3]{}

\acknowledgments
\textbf{Acknowledgements.--- } We thank María Cea, Mari Carmen Bañuls, Patrick Emonts, Hugues Meyer and Kay Brandner for fruitful discussions. Numerical simulations were performed using Mathematica, QuTiP~\cite{Johansson2013} and the ITensor library~\cite{Fishman2022}. We acknowledge funding from the Deutsche Forschungsgemeinschaft (DFG, German Research Foundation) through the Research Unit FOR 5413/1, Grant No. 465199066, through the Research Unit FOR 5522/1, Grant No. 499180199, and through the state of Baden-W\"urttemberg through bwHPC and the German Research Foundation (DFG) through grant no INST 40/575-1 FUGG (JUSTUS 2 cluster). We also acknowledge support from the Leverhulme Trust (Grant No. RPG-2024-112) J.d.B. and C.D.F. acknowledges support from the Alexander von Humboldt Foundation through a postdoctoral research fellowship. This work is supported by ERC grant OPEN-2QS (Grant No. 101164443, https://doi.org/10.3030/101164443).

\bibliography{biblio.bib}

\let\addcontentsline\oldaddcontentsline 

\clearpage
\onecolumngrid

\setcounter{equation}{0}
\setcounter{page}{1}

\setcounter{figure}{0}
\setcounter{table}{0}
\makeatletter
\renewcommand{\theequation}{S\arabic{equation}}
\renewcommand{\thefigure}{S\arabic{figure}}
\renewcommand{\thetable}{S\arabic{table}}
\setcounter{secnumdepth}{1}

\begin{center}
{\Large SUPPLEMENTAL MATERIAL}
\end{center}
\begin{center}
\vspace{0.8cm}
{\Large From Classical Stochastic to Monitored Quantum Dynamics: Dynamical Phase Coexistence in East Circuit Models}
\end{center}
\begin{center}
Marcel Cech,$^{1}$ Johan du Buisson,$^{1}$ Cecilia De Fazio,$^{1}$ Federico Carollo,$^{2,3}$ and Igor Lesanovsky$^{1,4}$
\end{center}
\begin{center}
$^1${\em Institut f\"ur Theoretische Physik and Center for Integrated Quantum Science and Technology, }\\
{\em Universit\"at T\"ubingen, Auf der Morgenstelle 14, 72076 T\"ubingen, Germany}\\
$^2${\em Dipartimento di Fisica, Sapienza Università di Roma, Piazzale Aldo Moro 2, 00185 Rome, Italy}\\
$^3${\em Centre for Fluid and Complex Systems, Coventry University, Coventry, CV1 2TT, United Kingdom}\\
$^4${\em School of Physics and Astronomy and Centre for the Mathematics}\\
{\em and Theoretical Physics of Quantum Non-Equilibrium Systems,}\\ 
{\em The University of Nottingham, Nottingham, NG7 2RD, United Kingdom}
\end{center}

\tableofcontents


\section{Average state dynamics and stationary state}
\label{sec:dynamicsandss}
In this section, we briefly discuss the average state dynamics of the monitored Floquet-quantum East model introduced in the main text. 
We begin by introducing the average state $\rho(t)$, i.e., obtained by averaging over all stochastic realizations at a fixed time $t$. This state evolves under the quantum map $\mathcal{E}$ according to
\begin{align}
    \rho(t+1) = \mathcal{E}[\rho(t)] = \sum_{\mathbf{k}} K_\mathbf{k} U \rho(t) U^\dagger K_\mathbf{k}^\dagger \, ,
    \label{eq:Kraus_map}
\end{align}
where $U$ describes the unitary evolution and $K_\mathbf{k}= \prod_i K_{k_i, i}$, with $K_{k_i, i}$ defined in  Eq.~\eqref{eq:weak_measurements}, represents measurements with outcomes $\mathbf{k} = (k_1, k_2, \dots, k_L)$. In particular, the operator $U= U_o U_e$ implements the Trotterized evolution over a single timestep displayed in Fig.~\ref{fig:fig1}(a) with 
\begin{equation}
   U_e = \bigotimes^{\lceil (L-1)/2 \rceil} _{j=0} u_{2j,2j+1}  \,, \quad U_o= \bigotimes^{\lfloor L/2 \rfloor} _{j=1} u_{2j-1,2j} \,,
\end{equation}
and the gate $u_{i-1,i}$ defined in Eq.~\eqref{eq:East_gate}. Note that open boundary conditions are implemented by $n_0 = 1$, implying that the qubit at $i=1$ undergoes an unconditioned rotation controlled by $\omega$ when $U_e$ is applied. This then implies that the Hilbert space does not have any disconnected subspaces for all rotation angles $\omega$ except when $\omega = 0\!\!\mod 2\pi$ where the dynamics is deterministic.\\

We can now argue that the stationary state is the structureless, infinite-temperature state. Given that the Kraus operators in Eq.~\eqref{eq:weak_measurements} are hermitian, the fully mixed state is a stationary state of the dynamics, i.e., $\rho_\mathrm{ss} = \mathds{1} / 2^L$ and $\mathcal{E}[\rho_\mathrm{ss}] = \rho_\mathrm{ss}$. Uniqueness of this stationary state is then established by noting that the Hilbert space is fully connected due to the boundary conditions discussed above. 

\section{Dynamical activity of space-time records } 
In this section, we introduce the dynamical activity, which serves as order parameter characterizing the dynamical phase transition in this work. Furthermore, we clarify how dynamical phase transitions can be studied in a similar way as equilibrium phase transitions using tools from dynamical large deviation theory. In particular, we treat individual space-time records as microstates featuring in a partition function, whose probabilities are weighted according to their activity. 

As described in the main text, the time-integrated activity for a single trajectory $\eta = [k_i(t)]_{i,t}$\,, is defined as
\begin{equation}
    A_{L,T}(\eta) = \sum_{i = 1}^L \sum_{t = 1}^T k_{i}(t)\,.
    \label{eq:dynamicalactivity}
\end{equation}
Here, $k_{i}(t)$ are the ancilla measurement outcomes recorded while monitoring the entire system of length $L$ for a time interval $T$. 
Recalling that the probability to observe a given measurement record $\eta$ is $\pi(\eta)$, the average activity density is computed as
\begin{equation}
    \label{eq:averag_activity_definition}
    a_{L,T}(0) = \frac{1}{LT} \sum_{\{\eta\}} \pi(\eta) A_{L, T}(\eta) \, .
\end{equation}
In the following, we now discuss both the value of the average activity density as well as its dynamical partition function, which allows us to study the full dynamical phase diagram. 

\subsection{Average activity density in the stationary state}
We now compute the average activity density, assuming that the system is initialized in the (trivial) stationary state (see $\rho_\mathrm{ss} = \mathds{1}/ 2^L$ in Sec.~\ref{sec:dynamicsandss}). In this case, the activity density $a_{L, T}$ does not depend on the time $T$~\cite{Landi2023,Cech2025}, and we have that
\begin{align}
    a_{L, T}(0) &= \frac{1}{L} \operatorname{Tr} \bigg\{ \sum_\mathbf{k} \bigg(\sum_i k_i \bigg) K_\mathbf{k} \underbrace{U \rho_\mathrm{ss} U^\dagger}_{=2^{-L}} K_\mathbf{k}^\dagger \bigg\} 
    = \frac{1}{L \, 2^L} \operatorname{Tr} \bigg\{ \sum_i \underbrace{\sum_\mathbf{k} k_i  K_\mathbf{k} K_\mathbf{k}^\dagger}_{(\sin^2\!\gamma) n_i} \bigg\} 
    = \frac{1}{L \, 2^L} \underbrace{\operatorname{Tr} \bigg\{\sum_i (\sin^2\!\gamma)\, n_i \bigg\}}_{= L \, 2^{L-1} \sin^2\!\gamma} 
    = \frac{\sin^2\!\gamma}{2} \, .
    \label{eq:average_activity}
\end{align}
We observe that $a_{L, T}(0)$ does not depend on $L$ or the unitary rotation angle $\omega$, but only on the measurement strength $\gamma$. Consequently, also the average activity density in the thermodynamic limit is given by $a(0) =\lim_{L,T\to \infty} a_{L, T} (0 )=\sin^2\!\gamma / 2$. In particular, it increases monotonically from $a(0) = 0$ for $\gamma = 0$, where no measurement of the system takes place, to $a(0) = 1/2$ for the projective measurements at $\gamma = \pi/2$. 
Note also that the result in Eq.~\eqref{eq:average_activity} can be directly understood in the following way: In the infinite temperature state, where $\bra{1_i}\rho_\mathrm{ss}\ket{1_i} = \bra{0_i}\rho_\mathrm{ss}\ket{0_i} = 1 / 2$, the probability to measure $k=1$ is $\sin^2\!\gamma / 2$ [cf.~Eq.~\eqref{eq:weak_measurements}]. 


\subsection{Dynamical partition function and large deviations}
\label{sec:largedeviation}
For the remaining part of this section, we now discuss in detail the dynamical partition function 
\begin{equation}
    \label{eq:dynamical_partition_function}
    \mathcal{Z}_{L,T}(s) = \sum_{\eta} \pi(\eta)e^{-s A_{L,T}(\eta)} \, ,
\end{equation}
with $\mathcal{Z}_{L,T}(0) = 1$, due to normalization. This object can be understood as a reweighting of the probabilities of trajectories according to a Boltzmann-like factor $e^{-s A_{L, T}(\eta)}$. In analogy with statistical physics, this can be interpreted as a canonical ensemble. We remind the reader that the dynamical partition function can develop non-analytic behavior in the thermodynamic limit $L, T \rightarrow \infty$, indicating that a  dynamical phase transition occurs. 
Furthermore, we note that Eq.~\eqref{eq:dynamical_partition_function_tilted_operator} can be rewritten as
\begin{equation}
    \label{eq:dynamical_partition_function_tilted_operator}
    \mathcal{Z}_{L,T}(s) = \mathrm{Tr}\left\{(\mathcal{E}_s)^T [ \rho_{ss} ]\right\} \, ,
\end{equation}
featuring the so-called tilted dynamical map $\mathcal{E}_s$. This superoperator is obtained from the original dynamical map $\mathcal{E}$ in Eq.~\eqref{eq:Kraus_map} by modifying the local Kraus operators according to $K_{k,i} \rightarrow e^{-sk/2}K_{k,i}$~\cite{Cilluffo2021}. Viewed in this way, the dynamics generated by $\mathcal{E}_s$ corresponds to a biased dynamics, where active trajectories are more ($s<0$) or less ($s>0$) likely than in the original ($s = 0$) dynamics~\cite{Carollo2018}. \\

Another important aspect of the dynamical partition function in Eq.~\eqref{eq:dynamical_partition_function} is that we can calculate the average activity density for arbitrary values of $s$ as stated in Eq.~\eqref{eq:activity_s_ensemble}, and equivalently as $a_{L,T}(s) = -\theta_{L, T}'(s)$ in terms of the so-called scaled cumulant-generating function (SCGF)~\cite{Garrahan2010}
\begin{align}
    \label{eq:SCGF_finite}
    \theta_{L, T}(s)  = \frac{1}{LT} \log \mathcal{Z}_{L, T}(s) \, .
\end{align}
This quantity plays an important role in the theory of large deviations~\cite{Touchette2009}, especially for the case that the following
\begin{equation} \label{eq:scgf}
    \theta(s) = \lim_{L,T\rightarrow \infty} \frac{1}{LT}\log \mathcal{Z}_{L,T}(s)
\end{equation}
exists, and is differentiable in $s$. Here, the probability distribution of $a = A_{L,T}/LT$ satisfies, for large $L$ and $T$, the asymptotic form
\begin{equation}
    P_{L,T}(a) \asymp e^{-LT \phi(a)}.
\end{equation}
In this case, the activity is said to satisfy a \textit{large-deviation principle} with rate function $\phi(a)$, which can be obtained from knowledge of $\theta$ in terms of a Legendre-Fenchel transformation~\cite{Touchette2009,Garrahan2010}. The meaning of the large deviation principle is that the dominant contribution to the distribution $P_{L,T}(a)$ is that of a decaying exponential in $LT$, with the rate of that decay controlled by the rate function $\phi(a)$. 
Again, the connection to statistical mechanics is apparent in noting that $\phi(a)$ plays the role of an entropy density. As such, corrections to this asymptotic form are sub-exponential in $LT$. 
Furthermore, even when $\theta(s)$ is not differentiable everywhere (as may be the case when a dynamical phase transition occurs), $\theta(s)$ can still be used to obtain the convex envelope of the rate function $\phi(a)$. \\

In practice, we often cannot determine the limit in Eq.~\eqref{eq:scgf}. In this case, since the $L \rightarrow \infty$ limit typically proves the more challenging of the two limits, we can sometimes calculate 
\begin{equation} \label{eq:SCGF_infiniteT}
    \theta_L(s) = \lim_{T \rightarrow \infty} \frac{1}{LT} \log \mathcal{Z}_{L,T}(s)
\end{equation}
as the dominant eigenvalue in a spectral problem~\cite{Cilluffo2021,Cech2025}. If this is not feasible, we must resort to numerically calculating $\theta_{L,T}(s)$ for large $L$ and $T$ according to Eq.~\eqref{eq:SCGF_finite}. 


\section{Exact results for the classical Floquet-East model}
\label{sec:classical_Floquet_East}
In this section, we now discuss the results for the classical stochastic Floquet-East model obtained for the special case of projective measurements, i.e., $\gamma = \pi/2$. 
First, we show explicitly that, for this case, the monitored quantum dynamics reduces to the classical stochastic Floquet-East model in Ref.~\cite{DeFazio2024}. Thereafter we utilize this observation to show that a dynamical phase transition occurs in the thermodynamic limit. Finally, we briefly describe how the activity density $a_L(s)$ shown in Fig.~\ref{fig:fig2} can be obtained via an eigenvalue calculation. 

\subsection{Correspondence to the classical stochastic Floquet-East model}
We start by recalling that, when $\gamma = \pi/2$, we can easily see that the Kraus operators given in Eq.~\eqref{eq:weak_measurements} become projection operators 
\begin{equation}
    K_{0,i} = \mathds{1}-n_i \,, \quad K_{1,i} = -i \, n_i \,,
\end{equation}
Consequently, the measurement of the ancilla qubits projectively measures all system qubits. This establishes a one-to-one correspondence between the states of system qubits and ancilla measurement outcomes: following the measurement of the ancillas, each system qubit is in one of the two computational basis states $\ket{0}$ (corresponding to measurement outcome $k = 0$) or $\ket{1}$ (for measurement outcome $k = 1$). Therefore, at the start and end of each timestep of the system's evolution, the system qubits are in a fully classical state, and the evolution of the system can be viewed as a probabilistic mapping of classical states to classical states. In particular, the system can be viewed as a Markov chain with transition matrix having the entries
\begin{equation}
    \label{eq:transition_matrix}
    (P_{\mathrm{cl}})_{mn} = \bra{m} \mathcal{E}\left[\,\ketbra{n}{n} \, \right]\ket{m},
\end{equation}
where $\ket{m}, \ket{n} \in \{\ket{0\dots00}, \ket{0\dots01}, \ldots,\ket{1\ldots11}\}$ here denote the $2^L$ possible $L$-qubit product states of the computational basis states. States are enumerated according to the binary representation of the integers $m,n$ which take values in $0, \ldots 2^L - 1$, so that $\ket{m} = \ket{\mathrm{bin}(m)}$, establishing a connection between matrix entries and multi-qubit states.

\begin{figure}
    \centering
    \includegraphics{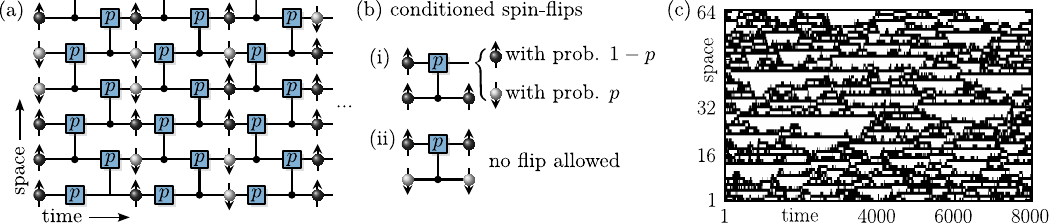}
    \caption{\textbf{Dynamical phase coexistence in the classical stochastic Floquet-East model.} (a)~Circuit representation of the stochastic dynamics. (b)~Stochastic East gates controlled by spin-flip probability $p$. (i)~spin-flip allowed; (ii)~ spin-flip forbidden. (c)~Stochastic realization. Black = spin up = 1; white = spin down = 0. The trajectory shows a typical stochastic realization starting from a random initial state for $L = 64$ spins for $p = \sin^2\omega$, with $\omega = 0.1$ as in the main text.} 
    \label{fig:fig2_SM}
\end{figure}

We can understand the Markovian stochastic dynamics further by utilizing the brickwork structure of the quantum circuit, shown in Fig.~\ref{fig:fig1}, in addition to the deferred measurement principle, which states that we can shift our projective measurements from the end of the timestep to a carefully selected intermediate stage of the timestep. This can be easily demonstrated, since the projective measurement commutes with the unitary gates (\ref{eq:East_gate})
\begin{equation}
    K_{k, i} e^{-i \omega n_{j-1} \sigma_{j}^x} = e^{-i \omega n_{j-1} \sigma_{j}^x}K_{k, i}
\end{equation}
for all $i \neq j$. Instead of applying measurements at the end of the timestep, we can therefore projectively measure qubit $i$ immediately following application of the gate $e^{-i \omega n_{i-1} \sigma_i^x}$. The effect of this procedure is illustrated in Fig.~\eqref{fig:fig2_SM}(b): qubit $i$ flips with probability $p = \sin^2\!\omega$ if qubit $i-1$ is in state $\ket{1}$ [panel (i)], and stays in its current state if $i-1$ is in state $\ket{0}$ [panel (ii)]. Our model therefore reduces to the kinetically constrained stochastic dynamics of a classical Floquet-East model~\cite{DeFazio2024}, obtained from the brickwork circuit illustrated in Fig.~\ref{fig:fig2_SM}(a). An example of typical stochastic realization of this dynamics is shown in Fig.~\ref{fig:fig2_SM}(c); we can see explicitly the coexistence of active and inactive clusters.

We note that for the special case where $\omega = \pi/2$ the probability $p$ of a spin flipping when the East constraint is satisfied is $p = 1$. The dynamics of our system then reduces to a \textit{deterministic} classical Floquet-East model~\cite{Klobas2024}. 

\subsection{Dynamical phase coexistence in the classical stochastic Floquet-East model}
\label{sec:stochasticEast}
Below, we show that the dynamical partition function defined in Eq.~\eqref{eq:dynamical_partition_function_tilted_operator} becomes non-analytic in the thermodynamic limit $L,T \to \infty$, and thus exhibits a discontinuity in its derivatives at $s = 0$ that can be associated with the presence of a phase transition. To this end, we follow a similar approach to Refs.~\cite{Klobas2024,DeFazio2024}. However, we stress that the activity considered here is a different observable: in particular, it counts the number of ancilla measurement outcomes (or, due to the one-to-one correspondence in the classical dynamics, the number of qubits) in state one rather than counting the number of flips or state changes. Furthermore, the dynamical partition function is here defined for the entire system size $L$ and total time $T$, and is therefore a global quantity rather than one defined on smaller space-time windows of the dynamics, as in the references above. 

We proceed by studying the behavior of $\mathcal{Z}_{L,T}(s)$ for the separate cases of $s \approx 0$ and $s = \infty$. For $s$ sufficiently close to $0$, we can write
\begin{equation}
    \mathcal{Z}_{L,T}(s) = 1 - s\langle A_{L,T}\rangle_0 + \mathcal{O}(s^2) = 1 - \frac{LT}{2}s + \mathcal{O}(s^2),
\end{equation}
where we have used the fact that $a_{L, T}(0) = \expval{A_{L, T}}_0/LT = (\sin^2\!\gamma)/2 = 1/2$ in this case. For $s = \infty$, we can use the fact that the exponential $e^{-s A_{L,T}}$ is now non-zero only for those trajectories that have $0$ total activity, so all spins must be down for all time-steps. There are only two such trajectories: the trajectory for which the system starts in the `all down' state, and the first spin never flips, and the trajectory in which the system is initialized in the state with all spins, except the first, in the down state, where the first spin is immediately flipped to the down state. Assuming that the system is initialized in the stationary states, where all computational basis states are equally likely, we then have that
\begin{equation} \label{eq:P_inactive_classical}
    \mathcal{Z}_{L,T}(s = \infty) = P_{L,T}(0) = 2^{-L}(1 - p)^T + 2^{-L}p(1-p)^{T-1} = 2^{-L}(1 - p)^{T-1},
\end{equation}
with $p = \sin^2\!\omega$, and where $P_{L,T}(0)$ was introduced previously. Importantly, we note for later use that $\log P_{L,T}(0)$ scales proportionally to the perimeter of the space-time region $L\times T$. This means that for $s \approx 0$, we have
\begin{align}
    \theta_{L,T}(s) &= \frac{1}{LT}\log \mathcal{Z}_{L,T}(s) \\
    &= -\frac{s}{2} + \mathcal{O}(s^2)
\end{align}
and 
\begin{equation} \label{eq:SCGF_s_infinite}
    \theta_{L,T}(s = \infty) = \frac{1} {LT} \log \mathcal{Z}_{L,T}(s = \infty) = - \frac{\log 2}{T} + \frac{(T-1)\log(1 - p)}{LT}.
\end{equation}
Importantly, as $L,T\rightarrow \infty$ we have that $\theta_{L,T}(s = \infty) \rightarrow \theta(s=\infty) = 0$. We also have that $\theta(0) = 0$ by definition of the scaled-cumulant generating function. Using the convexity of $\theta(s)$, we can therefore conclude that $\theta(s\geq0) = 0$, since $\theta(0) = 0$ and $\theta(s = \infty) = 0$. This means that in the thermodynamic limit, the SCGF is given for small negative $s$ by
\begin{equation}
    \theta(s) = -\frac{s}{2}
\end{equation}
and $\theta(s \geq 0) = 0$, so that $\theta$, and hence $\mathcal{Z}_{L,T}(s)$, is non-analytic (non-differentiable) at $s = 0$, proving the existence of a dynamical phase transition in the model for $\gamma = \pi/2$. 

From this analysis we can also see that, in the thermodynamic limit, the activity density $a(s) = \theta'(s)$, as defined in Eq.~\ref{eq:activity_s_ensemble}, becomes discontinuous at $s = 0.$ In particular, 
\begin{equation}
    a(0) = \frac{1}{2}, \quad a(0^+) = 0,
\end{equation}
showing that the typical dynamics at $s = 0$ exists on the coexistence line between active ($a(0) = 1/2$) and inactive ($a(0^+) = 0$) phases. 

Finally, we note that the above analysis breaks down for the deterministic case, with $p = 1$. This can be seen explicitly from (\ref{eq:SCGF_s_infinite}), which features the term $\log(1 - p)$ that becomes infinite when $p = 1$. The reason for this is that the probability of the completely inactive trajectory is zero, since the first spin will flip at every timestep. In the deterministic case, we also find that the model's state space is no longer fully connected. Depending solely on the initial state, only a certain subset of states is reached.
Due to this ergodicity-breaking, the special case of $p = 1$ must be treated with other methods, and we do not discuss it further here.

\subsection{Exact calculation of $a_L(s)$}
Here, we detail how to obtain the activity density $a_L(s)$ as shown in Fig.~\ref{fig:fig2}, which is found as $a_{L}(s) = -\theta'_L(s)$. It is known that the infinite-time SCGF $\theta_L(s)$ can be found by calculating the dominant eigenvalue in a spectral problem. In particular,
\begin{equation}
    \theta_L(s) = \frac{1}{L}\log \left(\mathrm{dom}\left(P_{\mathrm{cl}}(s) \right)\right),
\end{equation}
where the \textit{tilted matrix} $P_{\mathrm{cl}}(s)$ has entries given by
\begin{equation}
    (P_{\mathrm{cl}}(s))_{mn} = (P_{\mathrm{cl}})_{mn}e^{-s f(m)},
\end{equation}
where 
\begin{equation}
    f(m) = \textnormal{number of qubits in the up state in state} \, \ket{m} \, .
\end{equation}
Generally, the function $f$ depends on the dynamical observable that is being studied; we simply give the form of $f$ for the activity, our observable of interest~\cite{Touchette2009}. 

Using Mathematica, we can efficiently construct the tilted matrix ${P}_{\mathrm{cl}}(s)$ and obtain its dominant eigenvalue. The main limitation in this method is the fact that the dimension of the tilted matrix scales as $2^L \times 2^L$. Hence, we can only calculate the dominant eigenvalue analytically for intermediate values of $L$. The results are shown for various values of $L$ in Fig.~\ref{fig:fig2}, showing pronounced crossovers between active and inactive phases for small $s^* > 0$. 

It is useful to contrast the activity density $a_{L}(s)$ for the classical stochastic Floquet-East model with that obtained in a system of non-interacting (NI) spins, where each spin is flipped at each time-step with probability $p$, regardless of the states of its neighbors. Here, the SCGF is that of the $L=1$ case with 
\begin{equation}
    \theta_{\mathrm{NI}}(s) = \log \left(\frac{1}{2} e^{-s} \left[-p e^s + 
        \sqrt{(p - 1)^2 (e^s + 1)^2 + 4 (2 p - 1) e^s} - p + e^s + 1
    \right] \right).
\end{equation}
The corresponding activity density $a_{\mathrm{NI}}(s) = -\theta_{\mathrm{NI}}'(s)$ is then shown as dashed line in Fig.~\ref{fig:fig2}. Notably, we observe that no sharp crossover appears. This is the expected result, given that the coexistence of active and inactive phases in the classical stochastic Floquet-East model is a collective effect that arises due to the kinetic constraint. 


\section{Further discussion on the dynamical free energy}
In this section, we provide additional details on how the free-energy crossover discussed in the main text is extracted. We also explain how the emergence of perimeter-scaling at large cluster sizes signals the presence of a dynamical phase transition. Technical details on the tensor-network simulations used to compute the dynamical free energy can be found in Section~\ref{sec:TNmethods}.
\subsection{Crossover timescale from area- to perimeter-dominated scaling}
\begin{figure}[ht]
    \centering
    \includegraphics{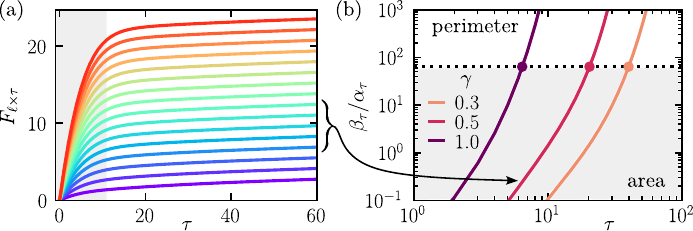}
    \caption{\textbf{Extraction of crossover time $\tau^*$.} (a)~Negative log-probability of inactive clusters. We show $F_{\ell \times \tau}$ for fixed $\ell$ (purple to red for $\ell = 2, 4, 6, ..., 32$) as a function of $\tau$ for $L = 64$, $\omega = 0.1$ and $\gamma = 0.5$. The gray shaded region guides the eye regarding the area- or perimeter-dominated scaling of $F_{\ell \times \tau}$. (b)~The ratio $ \beta_\tau /\alpha_\tau $  as a function of the temporal extent $\tau$ of the inactive cluster for several measurement strengths (colored lines). The parameters $\alpha_\tau$ and $\beta_\tau$ in Eq.~\eqref{eq:temporal_interface_tension}, shown here, are extracted by fitting the increment $\Delta F_{\ell,\tau}$ associated to clusters of fixed spatial size $\ell \in \{8, ..., 16\}$. The dashed line marks the threshold in Eq.~\eqref{eq:taustardef}, separating the area-scaling and perimeter-scaling regimes, where $L$ is the total system size. The dots indicate the extracted values of $\tau^*$ shown in Fig.~\ref{fig:fig4}(b).}
    \label{fig:fig4_SM_crossover}
\end{figure}

Here we analyze the scaling of the dynamical free energy $F_{\ell, \tau}$ in Eq.~\eqref{eq:freeenergy}. As illustrated in Fig.~\ref{fig:fig1}(c), this quantity is obtained by conditioning a local space-time region of the dynamics to contain no state-one outcomes in the ancilla records, thereby describing an inactive cluster of fixed size  $\ell \times \tau$ with $\ell \ll L $ and $\tau \ll T$.  To characterize the dependence of $F_{\ell, \tau}$ on the temporal extent $\tau$, we consider the increment 
\begin{equation}
    \Delta_\tau F_{\ell \times \tau} = F_{\ell \times (\tau + 1)} - F_{\ell \times \tau}\,,
\label{eq:temporal_interface_tension_def}
\end{equation}
that is, the change in free energy upon extending the cluster size by one timestep at fixed $\ell$. In particular, we employ the ansatz~\cite{Cech2025a}
\begin{align}
    \Delta_\tau F_{\ell \times \tau} = \alpha_\tau \, \ell + \beta_\tau \, ,
    \label{eq:temporal_interface_tension}
\end{align}
where $\alpha_\tau$ controls the area-contribution and $\beta_\tau$ the perimeter-contribution and both are obtained as fit parameters. The subscript $\tau$ indicates that we only consider the scaling in the temporal direction; a similar analysis could be performed for spatial variations. The ansatz above is motivated by the classical Floquet-East model~\cite{Klobas2024,DeFazio2024}, where close to a dynamical active-inactive phase transition, the free energy exhibits a crossover from area- to perimeter-scaling. A distinction between these two regimes can also be observed in the model considered here. In particular, in Fig.~\ref{fig:fig4_SM_crossover}(a), the free energy is plotted as a function of $\tau$ for several equidistant values of $\ell$. For small cluster sizes, we identify a linear growth with $\tau$, with a slope proportional to $\ell$, corresponding to a dominant area-scaling regime. However, for large cluster sizes, a clear linear growth with both $\tau$ and $\ell$ emerges, as indicated by the parallel lines, corresponding to a dominant perimeter-scaling regime. 

Furthermore the threshold in terms of the fit parameters $\alpha_\tau$ and $\beta_\tau$ can be estimated by identifying the perimeter-scaling regime with the following condition
\begin{equation}
    \label{eq:taustardef}
    \frac{\beta_\tau}{ \alpha_\tau} > L\,.
\end{equation}
The threshold on the right-hand side determines the dashed line in Fig.~\ref{fig:fig4_SM_crossover}(b), marking the point at which the perimeter contribution overtakes the area contribution for any potential values of $\ell$ in Eq.~\eqref{eq:temporal_interface_tension}. As such, 
the crossover time $\tau^*$ is then defined as the value of $\tau$ at which this condition starts to be satisfied. 
Although the definition of $\tau^*$ is not unique, this timescale offers an important insight on the measurement strength dependence of the dynamical phase coexistence of active and inactive space-time regions. In particular, as shown in Fig.~\ref{fig:fig4}(b), the crossover from area- to perimeter-dominated scaling is shifted to a larger $\tau$ as $\gamma$ decreases.

\subsection{Connection between inactive clusters and large deviation functions}
In this subsection, we connect the knowledge of dynamical free energy to the large deviation function introduced in Section~\ref{sec:largedeviation}. In particular, we argue that the area-to-perimeter crossover of the dynamical free energy (discussed in the previous subsection) should be interpreted as a finite-size manifestation of the underlying dynamical phase transition.

From the outset, we emphasize that the probability $p_{\ell,\tau}$  of finding an inactive space-time cluster in Eq.~\eqref{eq:freeenergy} can be defined as a local multi-point correlation function with support in a subsystem $\ell \ll L$. In contrast, large deviation functions such as  Eq.~\eqref{eq:activity_s_ensemble} are defined on large scales, typically the same as the system size $L$, where they can exhibit discontinuities in the thermodynamic limit. Despite operating at different scales, a connection between these two quantities can be established as we argue below. We also note that these arguments can be applied to trajectories from both classical stochastic and monitored quantum dynamics. 

We start by considering the statistics of the time-integrated activity in the space-time records. Suppose that we collect records of the entire system of size $L$ over a finite time $T$, and thus that we can compute the probability $P_{L,T}(a)$ of observing a given activity density  $a=A/(LT)$ within the space-time window  $L \times T$. 
If $ P_{L,T}(0) \neq 0$, we have that the zero-activity probability provides a lower bound 
\begin{equation}
\label{eq:lowerboundfinite}
 \theta_{L,T} (s) \geq - \frac{\log P_{L,T}\,(0)}{LT}
\end{equation}
where $s$ is the counting field that biases the activity~\cite{Garrahan2010}.
The lower bound resembles the free energy density  
\begin{equation}
\label{eq:densityinactive}
 \frac{F_{\ell \times \tau}}{\ell \tau} = -\frac{\log p_{\ell \times \tau}}{\ell \tau}  \,,
\end{equation}
where $F_{\ell \times \tau}$ is the dynamical free energy in Eq~\eqref{eq:freeenergy}. The important difference is that  Eq.~\eqref{eq:densityinactive} is computed within a (local) subregion $\ell \times \tau$ with $\ell \ll L$ and $\tau \ll T$, while the right-hand side of Eq.~\eqref{eq:lowerboundfinite} characterizes the minimal total activity over the full system. However, it is possible to show that these local and global quantities coincide in the thermodynamic limit if boundary effects become negligible. In particular, we define the large-deviation rate function 
\begin{equation}
I(a) = \lim_{L,T \to \infty} -\frac{\log P_{L,T}\,(a)}{LT}
\end{equation}
and expect that
\begin{equation}
    I(0)
    =
    \lim_{\ell,\tau \to \infty}
    \frac{F_{\ell \times \tau}}{\ell \tau},
    \label{eq:rate_function_from_free_energy}
\end{equation}
where the limit on the right-hand side  $\ell,\tau \to \infty$ is taken after $L,T \to \infty$ (see, e.g., Ref.~\cite{Klobas2024}), ensuring that boundary effects are negligible. 
Furthermore, in the thermodynamic limit, the lower bound gives
\begin{equation}
      \theta (s)  \geq - I(0) \,.
\end{equation}
As such, we expect this bound to be saturated as $s \to \infty$, which defines the asymptotic inactive branch
\begin{equation}
\label{eq:inactivebranch}
\theta^+=  \lim_{s\to\infty}  \theta (s) \sim -  \phi(0) \,.
\end{equation}
One way to look at the above limit is that we can arbitrarily increase the counting field $s$ to reach the most inactive dynamics. Recalling the activity density $a(s) =
- \theta^\prime (s)$, we have that 
\begin{equation}
\label{eq:inta}
   \int_{0}^s ds^\prime \,  a (s^\prime ) =-\theta (s)  \,.
\end{equation}
In particular, if the free energy $F_{\ell \times \tau}$ scales proportionally to the perimeter of the region for large but finite inactive clusters of size $\ell \times \tau$, Eq.~\eqref{eq:rate_function_from_free_energy} implies that $\phi(0)= 0$ in the thermodynamic limit.
Using Eq.~\eqref{eq:inactivebranch}, the normalization $\theta (0) = 0$ and convexity of the SCGF, it follows that $\theta (s) =0$ for all $ s\geq 0$. From the above equation, the activity density $a(s)$ therefore vanishes for positive $s$, and in particular $\lim_{s\to0^+}a(s) =0 $. However, Eq.~\eqref{eq:average_activity} gives $a(0) = (\sin^2 \gamma)/2$. Consequently, the activity density shows a discontinuity at $s=0$ in the thermodynamic limit. 

As we showed in Section~\ref{sec:stochasticEast}, the  perimeter-scaling in the first derivative of the activity density can be established exactly for $\gamma = \pi/2$, which proves the presence of a discontinuity in the activity density at $s=0$. More generally, as we have demonstrated above, the crossover to perimeter-dominated scaling observed in the dynamical free energy in Fig.~\ref{fig:fig4} can be interpreted as a finite-size precursor of the dynamical phase transition. 
 

\section{Perturbative extension of the classical stochastic Floquet-East Model}
\label{sec:perturbation}
In this section, we derive and discuss a perturbative extension of the classical stochastic Floquet-East model discussed in Sec.~\ref{sec:classical_Floquet_East}. The approach that we follow in this section closely resembles the derivation in Ref.~\cite{Lesanovsky2013a}. The central idea is that, for $\gamma \approx \pi / 2$, coherences are strongly suppressed such that we can treat small $\omega$ perturbatively. This perturbative approach sheds light on the scaling of the temporal extension of inactive regions that we utilized in Fig.~\ref{fig:fig4}(a). \\

To start, we note that the dynamics of the average state density matrix $\rho$ in Eq.~\eqref{eq:Kraus_map} can be split, for the aforementioned parameter regime, into fast dissipative dynamics and slow unitary dynamics $\mathcal{E} = \mathcal{E}_0 \circ \mathcal{E}_1$. The fast dynamics $\mathcal{E}_0[\rho] = \sum_\mathbf{k} K_\mathbf{k} \rho K_\mathbf{k}^\dagger$ are due to the strong monitoring, while the slow dynamics are unitary and given by $\mathcal{E}_1[\rho] = U \rho U^\dagger$. Considering only the fast dissipative dynamics with
\begin{equation}
    \begin{split}
        \mathcal{E}_0[\rho] &= \bigotimes_{i=1}^L \bigg( \sum_{k=0}^{1} K_{k, i} \, \rho \, K_{k, i}^\dagger \bigg) \\
        &= \bigotimes_{i=1}^L  
        \begin{pmatrix}
            \rho^{(i)}_{00} & \rho^{(i)}_{01} \cos{\gamma} \\
            \rho^{(i)}_{10} \cos{\gamma} & \rho^{(i)}_{11}
        \end{pmatrix}
        \, ,
    \end{split}
\end{equation}
where $\rho^{(i)}_{v w} = \bra{v_i} \rho \ket{w_i}$ denotes the $2^{L-1} \times 2^{L-1}$ dimensional matrix stored in $\rho$ after making $v, w\in \{0,1\}$ explicit~\cite{Lesanovsky2013}, we find that off-diagonal entries are altered by a factor of $\cos{\gamma}$ with every application of $\mathcal{E}_0$. For $\gamma \neq 0$ (or generally $\gamma \neq 0\!\!\mod \pi$), this implies that the stationary states of $\mathcal{E}_0$ are exactly given by the tensor product of the single-spin computational basis states $\{\ket{0}, \ket{1}\}$. Consequently, we define the projector $\mathcal{P}$ onto this manifold of $\mathcal{E}_0$ through
\begin{align}
    \label{eq:projector_relevant_subspace}
    \mu = \mathcal{P}\rho = \lim_{t \to \infty} \mathcal{E}_0^t[\rho] = \mathrm{diag}[\rho] \, .
\end{align}
For convenience, we denote the complementary projector by $\mathcal{Q} = \mathrm{Id} - \mathcal{P}$.
Note further that we can use a Baker-Campbell-Hausdorff identity to expand the evolution under the slow unitary dynamics as 
\begin{align}
    \label{eq:expansion_slow_dynamics}
    \mathcal{E}_1[\rho] = U \rho U^\dagger = \rho -i \mathcal{L}_1[\rho] - \frac{1}{2} \mathcal{L}_1\big[ \mathcal{L}_1[\rho]\big] + \mathcal{O}(\omega^3) \, ,
\end{align}
with $\mathcal{L}_1[\rho] = \omega \sum_{i = 1}^L \left[ n_{i-1} \sigma_i^x , \rho \right]$.
We later employ this expansion by noticing that applying $\mathcal{L}_1$ to a diagonal matrix (such as $\mu$) results in a matrix with only zeros on the diagonal. Formally denoted as $\mathcal{Q} \mathcal{L}_1\mathcal{P} = \mathcal{L}_1\mathcal{P}$, we can simplify $\mathcal{P}\mathcal{L}_1\mathcal{P} = 0$ and analogously $(\mathcal{Q} \mathcal{E}_0)^{t'} = \mathcal{Q} \mathcal{E}_0^{t'}$ by noticing that $\mathcal{E}_0$ leaves every (off-)diagonal term (off-)diagonal. Also, we note that $\mathcal{P}\mathcal{E}_0 = \mathcal{P}$ follows as a direct consequence from Eq.~\eqref{eq:projector_relevant_subspace}. From this point on, we omit square brackets unless necessary. \\

We now derive the effective dynamics in terms of a closed equation of motion for $\mu = \mathcal{P} \rho$. We hereby iteratively integrate out the dynamics of its complement $\mathcal{Q} \rho = (\mathrm{Id} - \mathcal{P}) \rho$. First, by utilizing $\rho(t+1) = \mathcal{E}\rho(t)$, we have that
\begin{align}
    \label{eq:dynamics_mu_iterative}
    \mu(t+1) = \mathcal{P} \mathcal{E}\rho(t) = \mathcal{P} \mathcal{E} \mu(t) + \mathcal{P} \mathcal{E} \mathcal{Q}\rho(t) \, ,
\end{align}
and
\begin{align}
    \label{eq:dynamics_Q_rho_iterative}
    \mathcal{Q}\rho(t) = \mathcal{Q} \mathcal{E}\rho(t-1) = \mathcal{Q} \mathcal{E} \mu(t-1) + \mathcal{Q} \mathcal{E} \mathcal{Q}\rho(t-1) \, .
\end{align}
Above, we have used $\mathcal{P} + \mathcal{Q} = \mathrm{Id}$ in order to separate the dynamics into the manifolds described by $\mathcal{P}$ and $\mathcal{Q}$, respectively. Here, the formal solution of Eq.~\eqref{eq:dynamics_Q_rho_iterative} is given by
\begin{align}
    \mathcal{Q}\rho(t) = \sum_{t'=1}^t (\mathcal{Q} \mathcal{E})^{t'} \mu(t-t') + (\mathcal{Q} \mathcal{E})^{t}\mathcal{Q}\rho(0) \, .
\end{align}
Substituting $\rho(0) = \mu(0)$ (and hence $\mathcal{Q}\rho(0) = 0$) into Eq.~\eqref{eq:dynamics_mu_iterative}, we obtain
\begin{align}
    \mu(t+1) &= \mathcal{P} \mathcal{E} \mu(t) + \mathcal{P} \mathcal{E} \sum_{t' = 1}^{t} (\mathcal{Q} \mathcal{E})^{t'} \mu(t - t') \, .
\end{align}
While this expression is exact, it involves the entire history of $\mu(t)$ and is therefore highly non-local. To overcome this obstacle, we perform a ``Markov approximation'' by (a)~replacing the time-non-local term $\mu(t-t')$ with $\mu(t)$ and (b)~send the summation limit $t$ to infinity in order to obtain a time-local and time-independent effective quantum channel of the form
\begin{align}
    \label{eq:time_local_effective_quantum_channel}
    \mu(t+1) = \mathcal{E}_\mathrm{eff}\mu(t) = \mathcal{P} \mathcal{E} \mu(t) + \mathcal{P} \mathcal{E} \sum_{t' = 1}^{\infty} (\mathcal{Q} \mathcal{E})^{t'} \mu(t)\, .
\end{align}
This approximation is justified given that the memory kernel $(\mathcal{Q} \mathcal{E})^t$ decays sufficiently fast with $t$, which is the case in the small $\omega$ regime. We now consider this expression term by term. We rewrite the first term as
\begin{align}
    \label{eq:first_term_time_local_effective_quantum_channel}
    \mathcal{P} \mathcal{E} \mu =\underbrace{\mathcal{P} \mathcal{E}_0}_\mathcal{=P} \mathcal{E}_1 \mu 
    \overset{\eqref{eq:expansion_slow_dynamics}}{=} \underbrace{\mathcal{P}\mu}_{= \mu} + \underbrace{\mathcal{P}\mathcal{L}_1\mu}_{=0} + \frac{1}{2} \mathcal{P}\mathcal{L}_1\mathcal{L}_1\mu + \mathcal{O}(\omega^3) \, .
\end{align}
Note that, in preparation to later combine the terms, we may use the identity $\mathcal{P} + \mathcal{Q} = \mathrm{Id}$ and write $\mathcal{P}\mathcal{L}_1\mathcal{L}_1\mu =\mathcal{P}\mathcal{L}_1 \mathcal{Q} \mathcal{L}_1\mu$ by invoking the fact that $\mathcal{P}\mathcal{L}_1\mathcal{P}\mathcal{L}_1\mu$ vanishes. For the other term in Eq.~\eqref{eq:time_local_effective_quantum_channel}, we find
\begin{align}
    \mathcal{P} \mathcal{E} \sum_{t' = 1}^{\infty} (\mathcal{Q} \mathcal{E})^{t'} \mu =& \underbrace{\mathcal{P} \mathcal{E}_0}_\mathcal{=P} \mathcal{E}_1 \sum_{t' = 1}^{\infty} (\mathcal{Q} \mathcal{E}_0 \mathcal{E}_1)^{t'} \mu
    \overset{\eqref{eq:expansion_slow_dynamics}}{=} \mathcal{P} \mathcal{L}_1 \sum_{t' = 1}^{\infty} \mathcal{Q} \mathcal{E}_0^{t'} \mathcal{L}_1 \mu + \mathcal{O}(\omega^3) \, ,
\end{align}
after using extensively the previously discussed identities.

We further simplify this expression using $\mathcal{Q} \mathcal{E}_0^{t'} \mathcal{L}_1 \mu = (\cos{\gamma})^{t'} \mathcal{Q} \mathcal{L}_1 \mu $, which holds due to the fact that $\mathcal{L}_1$ creates exactly a single coherence term for every spin that is affected by $\mathcal{E}_0^{t'}$. Subsequently, we evaluate the geometrical sum to obtain 
\begin{align}
    \label{eq:second_term_time_local_effective_quantum_channel}
    \mathcal{P} \mathcal{L}_1 \sum_{t' = 1}^{\infty} \mathcal{Q} \mathcal{E}_0^{t'} \mathcal{L}_1 \mu = \left( \frac{1}{1 - \cos{\gamma}} - 1 \right) \mathcal{P}\mathcal{L}_1 \mathcal{Q} \mathcal{L}_1\mu \, .
\end{align}
Finally, we can explicitly evaluate the term appearing in both Eq.~\eqref{eq:first_term_time_local_effective_quantum_channel} and Eq.~\eqref{eq:second_term_time_local_effective_quantum_channel} as
\begin{equation}
    \begin{split}
        \label{eq:simplified_expression}
    \mathcal{P}\mathcal{L}_1 \mathcal{Q} \mathcal{L}_1\mu = \mathcal{P}\mathcal{L}_1 \mathcal{L}_1\mu =& 2 \omega^2 \sum_{i, j = 1}^L \mathcal{P} \left( n_{i-1} \sigma_i^x \, \mu \,  n_{j-1} \sigma_j^x - \frac{1}{2} \{n_{i-1} \sigma_i^x n_{j-1} \sigma_j^x, \mu\}\right) \\
    =& 2 \omega^2 \sum_{i= 1}^L \left( n_{i-1} \sigma_i^x \, \mu \,  n_{i-1} \sigma_i^x - \frac{1}{2} \{n_{i-1}, \mu\}\right) \, ,
    \end{split}
\end{equation}
where we used that $\mathcal{P} \mathcal{L}_1$ returns the off-diagonal elements that were created by $\mathcal{Q} \mathcal{L}_1\mu$, back to the diagonal.

Combining the results of Eq.~\eqref{eq:first_term_time_local_effective_quantum_channel} and Eq.~\eqref{eq:second_term_time_local_effective_quantum_channel} according to Eq.~\eqref{eq:time_local_effective_quantum_channel}, and using the explicit form in Eq.~\eqref{eq:simplified_expression}, we obtain
\begin{align}
    \label{eq:eff_quantum_channel_East}
    \mathcal{E}_\mathrm{eff}\mu = \mu + \Tilde{\omega}^2 \sum_{i=1}^L  \left( n_{i-1} \sigma_i^x \, \mu \,  n_{i-1} \sigma_i^x - \frac{1}{2} \{n_{i-1}, \mu\}\right) + \mathcal{O}(\omega^3) \, ,
\end{align}
where we have defined $\Tilde{\omega}^2 = [1 +  (\frac{1}{1 - \cos\gamma} - 1)] \omega^2 = \cot^2{(\gamma/2)} \omega^2$. This effective generator can be directly translated into a transition matrix in the spirit of Eq.~\eqref{eq:transition_matrix} with
\begin{align}
    \label{eq:transition_matrix_East}
    P_\mathrm{cl} = \mathds{1} + \Tilde{\omega}^2 \sum_{i=1}^L n_{i-1} (\sigma_i^x - \mathds{1}) \, .
\end{align}
Here, two comments are in order. First, we notice that the transition matrix in Eq.~\eqref{eq:transition_matrix_East} is not guaranteed to be a physical stochastic matrix due to the eventual appearance of negative diagonal entries (i.e., for $L \Tilde{\omega}^2 > 1$). However, Eq.~\eqref{eq:transition_matrix_East} can also be understood as the first-order expansion of $e^{\mathcal{W}}$, where $\mathcal{W} = \Tilde{\omega}^2 \sum_{i=1}^L n_{i-1} (\sigma_i^x - \mathds{1})$ is equivalent to the Markov generator of the continuous-time East model~\cite{Jaeckle1991}. Consequently, the kinetic constraint is exactly preserved, and changes in the parameters only result in a modification of timescales. This modification is exactly the one that we also use in Fig.~\ref{fig:fig4}(a) to match the approximate crossover from area- to perimeter-dominated scaling of the inactive regions.


\section{Tensor network methods}
\label{sec:TNmethods}
In this section, we discuss the tensor network methods used to generate the data for our results in the main text that go beyond exact numerical calculations that were performed with QuTiP~\cite{Johansson2013}. On the one hand, this concerns simulating the pure state evolution of stochastic realizations and their corresponding space-time records as shown in Fig.~\ref{fig:fig1}(d). On the other hand, we discuss an appropriately constructed operator time-evolution to calculate the dynamical partition function $\mathcal{Z}_{L, T}(s)$ and the probability $p_{\ell \times \tau}$ of inactive space-time regions that enter in Figs.~\ref{fig:fig3} and~\ref{fig:fig4}, respectively. Our implementation~\cite{ZenodoData} leverages the open-source library \texttt{ITensor.jl}~\cite{Fishman2022} to perform the required tensor network contraction and truncations. All parameters that control the accuracy of these simulations are listed in Tab.~\ref{tab:TN_specifications}.

\begin{table}[b]
    \centering
    \caption{\textbf{Tensor network specifications.} We summarize the different tensor network methods as well as their hyper-parameters. List of abbreviations: SVD = Singular Value Decomposition, MPS = Matrix Product State, QJMC = Quantum-Jump Monte-Carlo, MPO-evo = Matrix Product Operator evolution. }
    \label{tab:TN_specifications}
    \setlength{\extrarowheight}{4pt}
    \begin{tabular}{c|c|c|c|c}
        Figure & Description & Method & maximal bond dimension & SVD cutoff $\epsilon$ \\
        \hline
        1 & Stochastic realization & MPS-QJMC & 256 (not saturated) & $10^{-10}$\\
        3 & Dynamical partition function & MPO-evo & 64 & $10^{-14}$ \\
        4 & Probability of inactive regions & MPO-evo & 32 & $10^{-14}$ \\
    \end{tabular}
\end{table}

\begin{figure}[ht]
    \centering
    \includegraphics{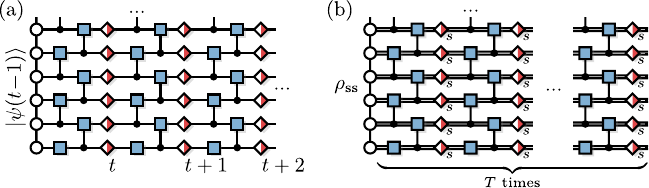}
    \caption{\textbf{Tensor network diagrams.} (a)~Stochastic realizations and space-time records. Using a combination of unitary gates (blue squares connected to their respective control qubit) and weak measurements (two-colored diamonds) described by the Kraus operators $K_{k, i}$ in Eq.~\eqref{eq:weak_measurements}, we can generate stochastic realizations $\ket{\psi(t)}$ alongside their corresponding space-time records. (b)~Dynamical partition function $\mathcal{Z}_{L, T}(s)$. Representing the stationary state $\rho_\mathrm{ss}$ as a vectorized MPO, we can compute the dynamical partition function by evolving this operator under both unitary evolution and modified Kraus operators (two-colored diamonds with subscript $s$) that include the counting field $s$.
    Note that a change in the local Kraus maps also allows us to simulate the probability $p_{\ell \times \tau}$ of inactive space-time regions. }
    \label{fig:fig_s1}
\end{figure}

\subsection{Stochastic realizations and space-time measurement records}
We simulate stochastic realizations and their corresponding space-time records $k_i(t)$ by closely following the quantum circuit structure shown in Fig.~\ref{fig:fig1}(a) in terms of a quantum-jump Monte-Carlo approach. Here, we employ a matrix product state (MPS) representation depicted in Fig.~\ref{fig:fig_s1}(a) to reach much larger system sizes compared to those accessible by means of exact diagonalization. Choosing a random product state in the computational basis as an initial state, the simulation of the stochastic dynamics always follows the same schematic for every discrete timestep. At first, we compute the evolution $\ket{\psi^-(t)} = U \ket{\psi(t-1)}$ before the monitoring via the time-evolving block decimation (TEBD) algorithm, where we make use of the East gates in Eq.~\eqref{eq:East_gate}. Subsequently, we simulate the effect and outcomes of the monitoring. Here, for the $i$th qubit, the probability of observing the outcome $k$ and evolving according to $\ket{\psi^-(t)} \to K_{k,i} \ket{\psi^-(t)} / \sqrt{\pi_k}$ is given by $\pi_k = \| K_{k,i} \ket{\psi^-(t)}\|^2$. Note that by sampling the vector $\mathbf{k}(t) = (k_1(t), ..., k_L(t))$ sequentially, this approach closely resembles the efficient sampling algorithm for MPS~\cite{Ferris2012} adapted to the case of positive operator-valued measures (POVMs) described by the set of Kraus operators $\{K_{k, i} \}$ in Eq.~\eqref{eq:weak_measurements}. This results in the state vector 
\begin{align}
    \ket{\psi(t)} = \frac{K_\mathbf{k} U \ket{\psi(t-1)}}{\| K_\mathbf{k} U \ket{\psi(t-1)} \|} \, ,
\end{align}
where $K_\mathbf{k}$ was introduced in Sec.~\ref{sec:dynamicsandss}. 
At all times, we control the accuracy of the MPS approximation by the cutoff value $\epsilon$ in the singular value decompositions (SVDs), while maintaining a maximal bond dimension of $\chi = 256$, which, in the case of Fig.~\ref{fig:fig1}(d), is not saturated even for $\epsilon = 10^{-10}$.

\subsection{Dynamical partition function}
We also use tensor network methods to calculate the dynamical partition function $\mathcal{Z}_{L, T}(s)$ by means of the operator time-evolution in Eq.~\eqref{eq:dynamical_partition_function_tilted_operator}. 
As mentioned before, this operator-evolution resembles the original average time-evolution $\rho(T) = \mathcal{E}^T[\rho(0)]$ in Eq.~\eqref{eq:Kraus_map}, upon modifying the local Kraus operators according to $K_{k, i} \to e^{-s k / 2} K_{k, i}$~\cite{Cilluffo2021}. 
Diagrammatically, this can be represented as done in Fig.~\ref{fig:fig_s1}(b), where diamonds with subscript $s$ denote the aforementioned modification of the local Kraus operators. Note also that double lines represent the folded space, where operators are represented as vectorized MPOs, and where we control the accuracy via the maximal bond dimensions $\chi$. \\

\begin{figure}[ht]
    \centering
    \includegraphics{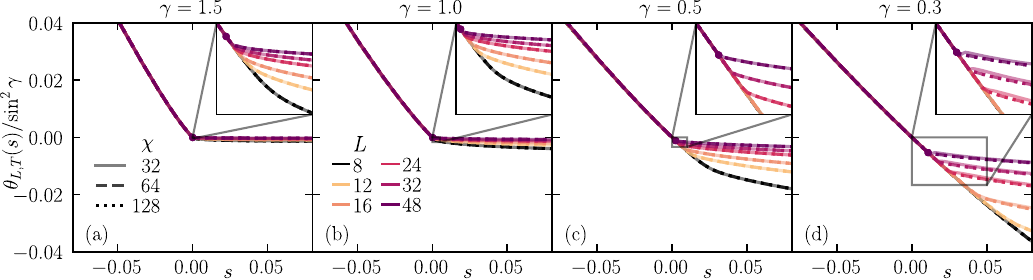}
    \caption{\textbf{Convergence checks in bond dimension for Fig.~\ref{fig:fig3}.} Panels show the scaled cumulant generating function $\theta_{L, T}(s)$ for different values of the maximal bond dimension $\chi$ (see linestyle and opacity) for different system sizes (see colors from black to purple with $L = 8$ to $L= 48$, respectively).  The inset shows a magnified view of the region for small $s\geq 0$. Throughout, we set $\omega = 0.1$ and simulate up to $T = 2\,000$ for $s < 0$ and $s \geq 0.1$, while considering $T = 10\,000$ for $s \in (0, 0.1)$ due to the increased interest in this region. Markers denote the exemplary simulations detailed in Fig.~\ref{fig:fig3_SM_time_evolution}.}
    \label{fig:fig3_SM}
\end{figure}

This approach allows for an efficient, yet well-controllable approximation of the dynamical phase diagram in Fig.~\ref{fig:fig3}. To validate its results, we now investigate the convergence of our simulations by comparing $\theta_{L, T}(s)$ [cf.~Eq.~\eqref{eq:SCGF_finite}] when increasing the accuracy by changing the maximal bond dimension $\chi$. In Fig.~\ref{fig:fig3_SM}, we observe that our simulations are well-converged given that small differences only appear in the comparison from $\chi = 32$ to $\chi = 64$ and $\chi = 128$ for $\gamma = 0.3$. \\

It is important to note that calculating $\mathcal{Z}_{L, T}(s)$ via the operator evolution in Eq.~\eqref{eq:dynamical_partition_function_tilted_operator} is, by definition, connected to the large deviation statistics of the activity at a finite time $T$ and finite system size $L$. 
Subsequently, both $\theta_{L, T}(s)$ in Eq.~\eqref{eq:SCGF_finite} and $a_{L, T}(s)$ in Eq.~\eqref{eq:activity_s_ensemble} correspond to the same finite time dynamics, where the initial state is described by an infinite temperature state. 
In order to quantify the influence of finite simulation times, we investigate the convergence of $\theta_{L, T}(s)$ by defining the quantity 
\begin{align}
    \lambda_{L, t}(s) = \frac{1}{L} \log \left(\frac{\mathcal{Z}_{L, t}(s)}{\mathcal{Z}_{L, t-1}(s)} \right)\, ,
    \label{eq:instantaneous_increment}
\end{align}
with $\mathcal{Z}_{L, 0}(s) = 1$. Consequently, $\lambda_{L, t}(s)$ is related to the instantaneous increment in calculating the SCGF in Eq.~\eqref{eq:SCGF_finite} via $\theta_{L, T} = \frac{1}{T} \sum_{t = 1}^{T} \lambda_{L, t}(s)$. In particular, the quantity in Eq.~\eqref{eq:instantaneous_increment} can be understood as the power-method approximation for the SCGF at stationarity (i.e., infinite times) due to the fact that $\theta_L(s)$ is connected to the dominant eigenvalue of $\mathcal{E}_s$\cite{Cilluffo2021,Cech2025}. 

\begin{figure}[ht]
    \centering
    \includegraphics{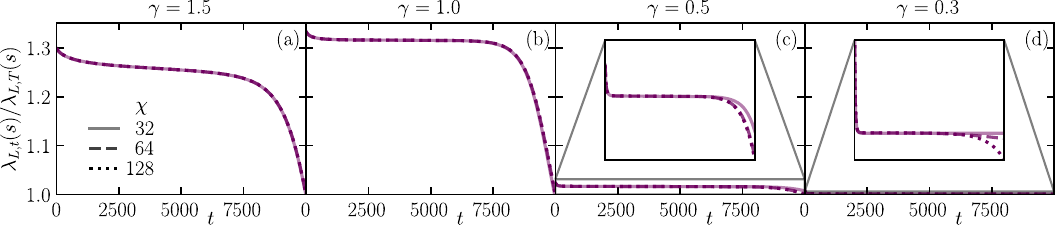}
    \caption{\textbf{Influence of finite simulation times.} We show the fraction of $\lambda_{L, t}(s)$ over the last increment $\lambda_{L, T}(s)$ that we have inferred for $T = 10\,000$ and $\chi = 128$. Different panels show different measurement strengths $\gamma$ for which we select representative values of $s$ that are close to their corresponding values of $s^*$ (see Fig.~\ref{fig:fig3} and Fig.~\ref{fig:fig3_SM}).  Throughout, we consider $L = 48$ and $\omega = 0.1$. The insets show the same data magnified. }
    \label{fig:fig3_SM_time_evolution}
\end{figure}

In Fig.~\ref{fig:fig3_SM_time_evolution}, we now consider $\lambda_{L, t}(s)$ [divided by its final value $\lambda_{L, T}(s)$ in the simulation to reduce the overall difference between calculations for different $\gamma$] for representative values of $s$ that are close to the corresponding crossover points $s^*$. 
While convergence to the dominant eigenvalue of $\mathcal{E}_s$ would be indicated by $\lambda_{L, t}(s) / \lambda_{L, T}(s) \to 1$ already for $t \ll T$, we observe that $\lambda_{L, t}(s)$ exhibits a metastable plateau as well as non-stationary behavior at late times. This observation is consistent with the expectation of a closing spectral gap as one approaches a dynamical phase transition. We address this point by noting that our simulations underestimate the sharpness of the crossover due to the fact that our finite time calculations are upper bounded by the dominant eigenvalue of $\mathcal{E}_s$ that is connected to $\theta_L(s)$ at infinite times. 

\subsection{Probability of inactive space-time regions}
Finally, we turn to the probability $p_{\ell \times \tau}$ of inactive space-time clusters. We again use an operator evolution similar to the one in Fig.~\ref{fig:fig_s1}(b) to calculate this quantity. 
In particular, we employ the combination of unitary dynamics described by $U$ and local monitoring according to $K_{0, i} \rho K_{0, i}^\dagger$ for sites $i$ that are contained in the space-time region $\ell \times \tau$ or $\sum_k K_{k, i} \rho K_{k, i}^\dagger$ for $i$ that lie outside this cluster, respectively. Distinguishing between sites in this manner implements the conditioning on an inactive space-time region such that the trace over the time-evolved operator after $\tau$ such timesteps accesses $p_{\ell \times \tau}$~\cite{Cech2025a}. 

In practice, we would also like to minimize the influence of the boundary in the spatial direction. Given that the East constraint and the open-boundary conditions introduce a directionality in the dynamics, where the $i$th site facilitates the dynamics of its neighbor at site $i+1$, we can achieve this by placing the $\ell \times \tau$-sized inactive region at $i \in \{L - \ell - 1, ..., L\}$. Additionally, we note that the influence of the boundary generally decays quickly with respect to the distance from the boundary at $i = 0$. This implies that one can evaluate inactive clusters from almost the whole space-time volume of individual trajectories in order to obtain more samples reducing statistical uncertainties.

\begin{figure}[ht]
    \centering
    \includegraphics{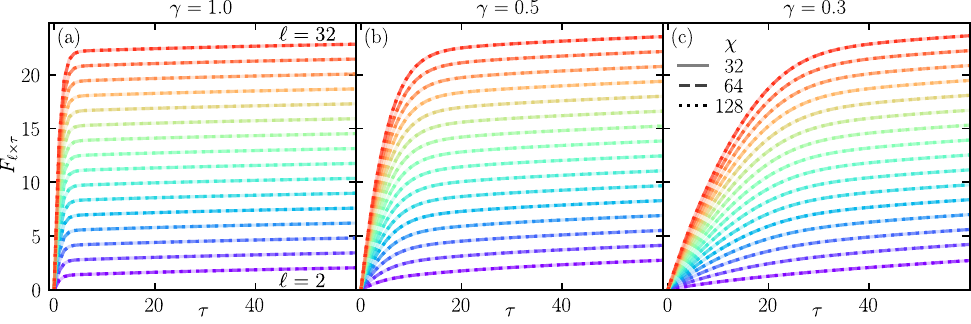}
    \caption{\textbf{Convergence checks for Fig.~\ref{fig:fig4}.} We show the negative log-probability $F_{\ell \times \tau}$ of inactive space-time regions when varying their temporal extension $\tau$ for $L = 64$, $\omega = 0.1$ and different measurement strengths $\gamma$. Colors from blue to red indicate the fixed spatial extensions $\ell \in \{2, 4, ..., 32\}$, while linestyle and opacity indicate the maximal bond dimension $\chi$. }
    \label{fig:fig4_SM}
\end{figure}

To validate the tensor network simulations for the probability $p_{\ell \times \tau}$ of inactive space-time regions, we now investigate the convergence of the corresponding dynamical free energy $F_{\ell \times \tau} = -\log p_{\ell \times \tau}$ with respect to the maximal bond dimension $\chi$ in Fig.~\ref{fig:fig4_SM}. The results show convergence with respect to $\chi$ for all considered measurement strengths $\gamma$. Note also that we have not rescaled the time axis, and again observe that the area-to-perimeter crossover occurs at larger values of $\tau$ as $\gamma$ decreases. \\

\end{document}